\input{aipcheck}

\documentclass[
    ,final            
  ]
  {aipproc}

\layoutstyle{8x11single}

\begin{document}

\title{European underground laboratories: An overview.}


\author{Lino Miramonti}{
  address={Physics Dept. of Milano University and 
National Institute of Nuclear Physics\\
via Celoria 16, 20133 Milano, Italy}
}

\begin{abstract}
Underground laboratories are complementary to those where the research in fundamental physics is made using accelerators.
This report focus on the logistic and on the background features of the most relevant laboratories in Europe, stressing also on the low background facilities available. 
In particular the report is focus on the laboratories involved in the new Europeean project ILIAS with the aim to support the European large infrastructures operating in the astroparticle physics sector.
\end{abstract}

\maketitle


\section{Introduction}
Thanks to the underground facilities one reaches the low radioactivity background environment needed to search for rare events in order to study the fundamental laws of nature such as those governing the proton stability or the nature of the neutrino. In addition the study of astroparticle physics phenomena such as neutrino from astrophysical objects and dark matter searches are also possible.
The underground laboratories are complementary to those where the research in fundamental physics is made using accelerators.
In these last decades different underground laboratories have been excavated all over the world; they differ in size, in the support of technology available and in the way to access them. In Europe the most relevant laboratories (not in shallow site) are the Italian Laboratori Nazionali del Gran Sasso, the French Laboratoire Souterrain de Modane, the Spanish Laboratorio Subterraneo de Canfranc, the UK Boulby Underground Laboratory, the Centre for Underground Physics in Pyh\"asalmi in Finland, the Ukranian Solotvina Underground Laboratory and the Russian Baksan Underground Laboratory. Table \ref{Compilation of European laboratories} summarizes the European underground laboratories.
In the next chapters we will concentrate in particular on the first four laboratories, involved in the new Europeean project ILIAS with the aim to support the European large infrastructures operating in the astroparticle physics area.\\
The most relevant sources of background are the penetrating muons, the primordial radionuclides in the rock and in all material constituting the detector and/or its shielding and the radon gas.\\
In this report we are focusing on the logistic and on the background features of the labs, stressing also on the low background facilities available.

\section{The Gran Sasso National Laboratory - LNGS}
The Gran Sasso National Laboratory (Laboratori Nazionali del Gran Sasso - LNGS) is the world's largest underground research center; it is financed by the Italian National Institute for Nuclear Physics (INFN), which coordinates and finances research in nuclear, subnuclear and elementary particle physics. The excavation and construction of the underground halls began in mid eighties. 
The laboratory is located in the highway tunnel between Teramo and L'Aquila under the "monte Aquila" (Gran Sasso mountain) in the center of Italy. Three main experimental halls (20 m high, 18 m wide and 100 m long) host big scientific experiments. Including the connecting tunnels and emergency passageways, the total volume is 180000 $m^{3}$ and the area is 13500 $m^{2}$. Because of the great quantity of water inside the Gran Sasso mountain, the natural temperature is about 6-7 $^{0}C$ and the humidity is approximately 100\%. 
In order to host scientific facilities, the experimental halls are waterproofed and heated; furthermore, a ventilation system provides about 35000 $m^{3}$ of air per hour from outside. The underground laboratory is connected to the external laboratory (situated about 8 km away) by means of fiber optical cables; the external structures, located in Assergi, include offices, laboratories, machine shop, seminar halls and a computer center for a total surface of about 12000 $m^{2}$.\\
The experimental activities presently ongoing at LNGS include all major research topics in the field of underground science: neutrino astrophysics (solar neutrinos (Borexino) and supernova neutrinos (LVD)), long baseline neutrino detection (CNGS project with the OPERA and ICARUS experiments), search for neutrino mass in neutrinoless $\beta \beta$ decay (Cuoricino/Cuore and Gerda), dark matter search (Dama, Libra, CRESST), nuclear astrophysics (LUNA), and several projects in the fields of geophysics, biology, and environmental sciences.

\begin{figure}
  \includegraphics[height=.3\textheight]{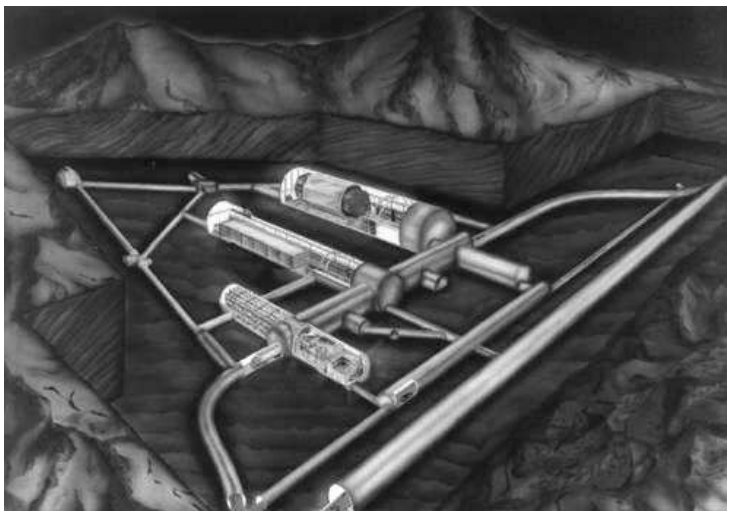}
  \caption{The Gran Sasso National Laboratory}
\label{LNGS}
\end{figure}

\subsection{Background characteristics and facilities}
The overburden rock is on the average about 1400 m, equivalent to about 3800 meter of water. The muon flux is reduced of about 6 orders of magnitude to a value of approximately 1.1 muons per square meter per hour, whereas the neutron flux is on the order of $10^{-6}$ neutrons per square centimeter per second depending on energy (see table \ref{neutron flux}).
The rock of the Gran Sasso mountain has a density of 2.71 $\pm$ 0.05 $g \cdot cm^{-3}$, and consists mainly of $CaCO_{3}$ and $MgCO_{3}$ \cite{Catalano1986}. Because of the presence of a particular type of rock (roccia nera marmosa), the content of uranium and thorium in hall A rock is about 10 to 30 times higher than in hall C. The primordial radionuclide content of the rock and of the concrete employed to cover the three big halls are reported in table \ref{primordial radionuclides}.
The low-level radioactivity topics carried out in the LNGS are: 
material selection and sample measurements (mainly with HPGe detectors);
radon groundwater monitoring (Environmental and geophysical monitoring of the Gran Sasso aquifer);
development and characterization of new detectors  (for nuclear spectrometry of environmental radioactivity).
At present, a 32 $m^{2}$ room is dedicated to low-level radioactivity measurements (for the future a new building with ca. 80  $m^{2}$ distributed on three floors is foreseen in hall A). This facility consists currently in nine high purity germanium detectors; in table \ref{table HPGe at LNGS} are reported the performances of the HPGe detectors at LNGS.

\begin{table}

\begin{tabular}{llcrc}
\hline
  & \tablehead{1}{r}{b}{Laboratory}
  & \tablehead{1}{r}{b}{Location}
  & \tablehead{1}{r}{b}{Overburden\\$[mwe]$}
  & \tablehead{1}{r}{b}{muon flux\\$[\mu$ $m^{-2}$ $h^{-1}]$} \\
\hline
& LNGS 	& Italy       & 3800   & 1.1   \\
& LSM   & France  		& 4800   & 0.17  \\
& LSC   & Spain  		  & 2450   & 7.2   \\
& IUS   & UK  	      & 2800   & 1.5   \\
& CUPP  & Finland     & 4000   & -     \\
& SUL   & Ukraina  	  &  430 	 & 6.2   \\

\hline
\end{tabular}
\caption{European underground laboratories. For the LNGS, LSM and LSC the overburden is the approximate mean depth, while for the IUS, CUPP and SUL the overburden is the vertical depth under flat surface.}
\label{Compilation of European laboratories}
\end{table}

\begin{table}
\begin{tabular}{lrrrr}
\hline
  & \tablehead{1}{c}{b}{40-2700 keV}
  & \tablehead{1}{c}{b}{352 keV}
  & \tablehead{1}{c}{b}{583 keV}
  & \tablehead{1}{c}{b}{1461 keV}   \\
\hline
GeMi     	&  611        & 5.6        & 2.1         & 5.2   \\
GePV     	&  482        & 2.8        & 2.1         & 3.2   \\
GeOr     	&  469        & 2.4        & 0.76        & 4.3   \\
GePaolo   &  266        & 0.83   		 & 0.38        & 1.4   \\
GeCris   	&   87        & $<$0.39    & $<$0.29     & 1.0   \\
GeMPI   	&   30        & $<$0.20    & $<$0.15     & 0.36   \\
\hline
\end{tabular}
\caption{Performances of the HPGe detectors at LNGS (Data @ 2002). Total and peak background count rate in $d^{-1}$ $kg^{-1}_{Ge}$. \cite{Laubenstein2005}}
\label{table HPGe at LNGS}
\end{table}

\section{The Modane Underground Laboratory - LSM}
The Modane Underground Laboratory (Laboratoire Souterrain de Modane - LSM) belongs to the French institutions of IN2P3-CNRS
(Centre National de la Recherche Scientifique) and DSM-CEA (Commissariat \`a l'Energie Atomique). It is located in the highway tunnel connecting Italy to France at about 6.5 km from both entrances under the "pointe du Fr\'ejus" at 1200 m over the sea level. This underground facility is also know as Fr\'ejus underground laboratory.
The underground laboratory consists of one main hall (11 m high, 10 m wide and 30 m long) and three secondary halls: a gamma spectroscopy hall of 70 $m^{2}$ and two little rooms of 21 $m^{2}$ and 18 $m^{2}$. The laboratory was created at the beginning of eighties in order to host a large calorimeter for  proton decay searches. 
The natural temperature of the rock is of the order of 28 $^{0}C$, but an air conditioning system, in addition to the ventilation system, reduces the temperature at about 20 $^{0}C$ with a relative humidity under 50\%. 
The external buildings are located in the Modane village (French side), and are composed of offices and workshops.\\
The present scientific programme of LSM include direct dark matter detection (Edelweiss), search for neutrinoless double beta decay (NEMO), and study of super-heavy nulei (SHIN). Several projects requiring ultra-low
levels of background in the field of environmental and geophysical sciences are also ongoing.

\begin{figure}
  \includegraphics[height=.3\textheight]{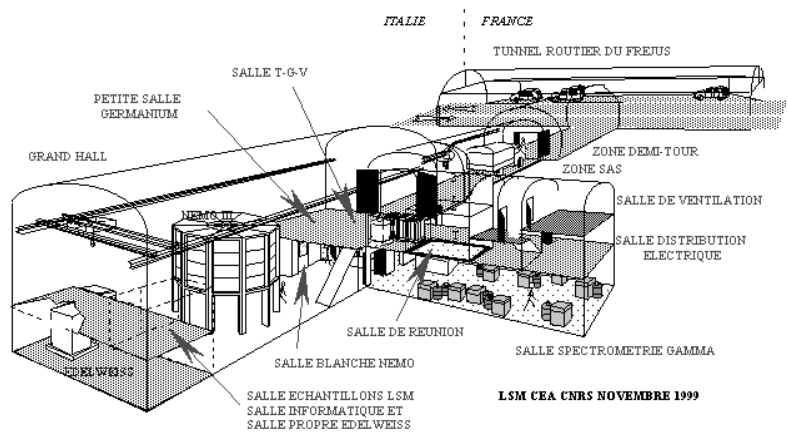}
  \caption{The Modane Underground Laboratory}
\label{LSM}
\end{figure}

\subsection{Background characteristics and facilities}
The rock overburden, essentially made of schist (with a density of about 2.73 $g \cdot cm^{-3}$) is about 1700 m, (corresponding to about 4800 mwe). The residual muon flux at the laboratory depth is of about 0.17 muons per square meter per hour. The neutron flux is $1.6 \cdot 10^{-6}$ $n$ $cm^{-2}$ $s^{-1}$ for neutrons having energy up to 0.63 eV and $4 \cdot 10^{-6}$ $n$ $cm^{-2}$ $s^{-1}$ for neutrons with energy ranging between 2 MeV and 6 MeV (see table \ref{neutron flux}).
In table \ref{primordial radionuclides} are reported the primordial radionuclide content of the rock and of the concrete.
Thirteen HPGe from 6 different laboratories of CNRS and CEA are available at LSM for material selection in fundamental physics searches and for samples measurement for environmental control, earth science, archaeology, biology and dating measurement.

\begin{table}
\begin{tabular}{lrrrr}
\hline
  & \tablehead{1}{c}{b}{$^{238}U$}
  & \tablehead{1}{c}{b}{$^{232}Th$}
  & \tablehead{1}{c}{b}{$K$}
  & \tablehead{1}{c}{b}{Ref.}   \\
\hline
LNGS Hall A (rock)       			& 6.80 $\pm$ 0.67 ppm   		& 2.167 $\pm$ 0.074 ppm	   & 160 ppm   & \cite{Wulandari2004} \\
LNGS Hall B (rock)       		 	& 0.42 $\pm$ 0.10 ppm   		& 0.062 $\pm$ 0.020 ppm	   & 160 ppm   & \cite{Wulandari2004} \\
LNGS Hall C (rock)       			& 0.66 $\pm$ 0.14 ppm   		& 0.066 $\pm$ 0.025 ppm	   & 160 ppm   & \cite{Wulandari2004}\\
LNGS all Halls (concrete)     & 1.05 $\pm$ 0.12 ppm   		& 0.656 $\pm$ 0.028 ppm	   &  -        & \cite{Bellini1991}\\
\hline
LSM (rock)       							& 0.84 $\pm$ 0.2 ppm   		  & 2.45 $\pm$ 0.2 ppm	     & 213 $\pm$ 30 $Bq \cdot kg^{-1}$   & \cite{chazal1998}  \\
LSM (concrete)       					& 1.9  $\pm$ 0.2 ppm   	  	& 1.4  $\pm$ 0.2 ppm	     & 77  $\pm$ 13 $Bq \cdot kg^{-1}$   & \cite{chazal1998}  \\
\hline
LSC             				& $2.2 \cdot 10^{-2}$ $\gamma$ $cm^{-2}$ $s^{-1}$   & 	  &     & \cite{Morales2004} \\
\hline
IUS            				  			& 70 ppb 		    & 	 125 ppb   & 1130 ppm  &							  \cite{Vitaly2004}\\
\hline
CUPP       	          	& 27.8-44.5 $Bq \cdot m^{-3}$   & 4.0-18.7  $Bq \cdot m^{-3}$	   & 267-625 $Bq \cdot m^{-3}$ ($^{40}K$) & 			\cite{Wang2000}  \\

\hline
\end{tabular}
\caption{Primordial radionuclides content in rock and concrete in the European underground facilities.}
\label{primordial radionuclides}
\end{table}

\section{The Canfranc Underground Laboratory - LSC}
The Canfranc Underground Laboratory (Laboratorio Subterraneo de Canfranc - LSC) is located in the now closed to traffic railway tunnel of Somport (6.9 km long) connecting Spain to France in the Aragonese Pyrenees, at 1080 m above the sea level. The operating institution is the Zaragoza University.
The underground facilities consist of one main hall called Lab3 (4.5 m high, 5 m wide and 20 m long) and a small laboratory called Lab1 formed by two small rooms of 18 $m^{2}$ each. A third laboratory, called Lab2 and now dismounted, consisting in a mobile pre-fabricated hut of about 20 $m^{2}$, is placed on the railway tracks and moved along the tunnel to operate at different overburden.
The natural temperature of the laboratory is 7 $^{0}C$ and the relative humidity is about 90\%. There is a constant flow wind coming from the French side which keeps the level of radon fairly low. The main laboratory is ventilated every four hours with fresh air at a rate of 1500 $m^{3}$ per hour. The air conditioning system provides average temperature of 21 $^{0}C$ and a relative humidity of about 60\%.
At the railway entrance (Spanish side) there is a building of the Spanish National Railway Company at disposal of the Zaragoza University for support of the activities. A pre-fabricated hut of about 20 $m^{2}$ is equipped for logistic link and monitoring.\\
The laboratory is presently being substantially enlarged both in size and logistic support (see below). Activities in the old lab included $\beta \beta$ decay search (IGEX-$\beta \beta$, Gedeon), and dark matter searches (Igex-DM, Rosebud); the scientific programme of the new laboratory is still under discussion.

\begin{figure}
  \includegraphics[height=.3\textheight]{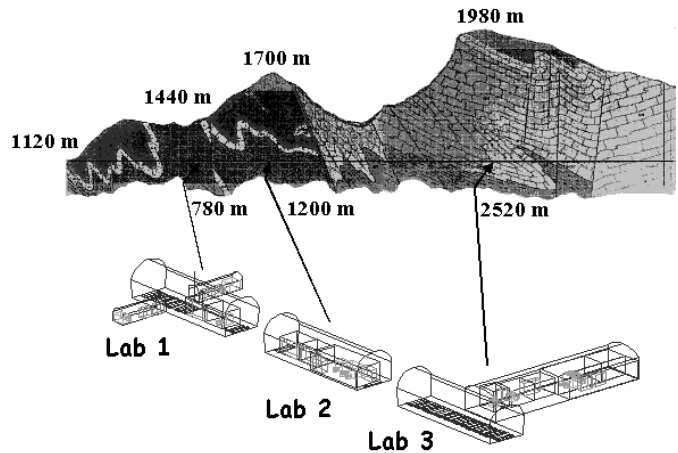}
  \caption{The Canfran Underground Laboratory}
\label{LSC}
\end{figure}

\subsection{Background characteristics and facilities}
Lab3 is located the Tobazo's peak (1980 m high) under an overburden of about 900 m of rock (limestone, mainly calcium carbonate), which correspond to about 2450 mwe; (the overburden of Lab1 is about 200 m of rock which correspond to about 675 mwe). 
Lab1 is used now only to store detectors and other materials (which  are so kept shielded from the cosmic radiation).
The muon flux in the main hall is 7.2 muons per square meter per hour.
The integrated neutrons flux from radioactivity is $3.82 \cdot 10^{-6}$ $n$ $cm^{-2}$ $s^{-1}$, whereas the integrated neutron flux from muon-induced is   $1.73 \cdot 10^{-9}$ $n$ $cm^{-2}$ $s^{-1}$ (see table \ref{neutron flux}).
The measured gamma flux inside the laboratory is about $2 \cdot 10^{-2}$ $\gamma$ $cm^{-2}$ $s^{-1}$.
An installation (AMBAR) for measuring low contents of radioactive contaminants in materials intended for low-background experiments is available.

\subsection{The new Canfranc Underground Laboratory}
Civil works are underway (50 m from the old facility) for the construction of the new Canfranc Underground Laboratory. It will consist in two experimental halls: the main one, oriented towards CERN is 11 m high, 15 m wide and 40 m long for fundamental physics research, and a second one (8 m high, 10 m wide and 15 m long) as low background facility. An access corridor housing offices, workshops and clean rooms for a total surface of about 1000 $m^{2}$ completes the new laboratory.

\section{The Boulby Mine Laboratory - IUS}
Unlike the previous laboratory, the IUS (Institute for Underground Physics) laboratory is not located into a tunnel but in a  salt mine near Sheffield, in northern England. The Boulby mine laboratory started in 1988 for the search for non-baryonic dark matter candidates.  
It consists of different areas: a first room called \textsl{Stub 2}, excavated in 1988, of about 300 $m^{2}$, a second room called \textsl{Stub 2a}, excavated in 1995, with an area of about 150 $m^{2}$, an area called \textsl{H area}, excavated in 1998, with an area of about 900 $m^{2}$, and the new \textsl{JIF area}, excavated in 2003, with a total area of about 2500 $m^{2}$.
The relative humidity is approximately 37\% with an environment radon concentration of about 5 $Bq \cdot m^{-3}$. The temperature of the surrounding rock is about 35 $^{0}C$, but in the experimental areas it is lowered to 28 $^{0}C$ thanks to an air conditioned system. 
Surface facilities, on the mine site, consisting in administration offices, computer facilities, conference room, workshop and storage space are available.\\
Research topics at Boulby are mostly concentrated on Dark Matter search with the Zeplin and Drift projects. 

\begin{figure}
  \includegraphics[height=.3\textheight]{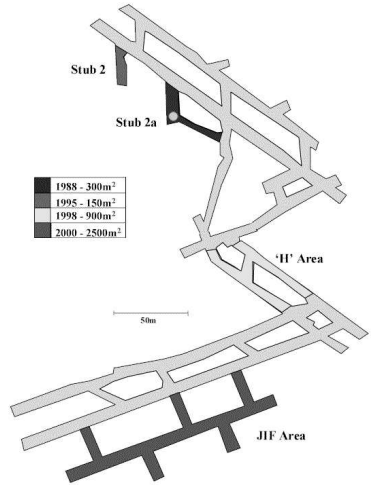}
  \caption{The Institute for Underground Science - Boulby Mine}
\label{IUS}
\end{figure}

\subsection{Background characteristics and facilities}
The depth of the laboratory is about 1070 m, corresponding to approximately 2800 mwe; at this depth the muon flux is about 1.5 muons per square meter per hour. The calculated neutron flux is about $2 \cdot 10^{-6}$ $n$ $cm^{-2}$ $s^{-1}$ for neutrons having energy greater than 100 keV, and about $1 \cdot 10^{-6}$ $n$ $cm^{-2}$ $s^{-1}$ for neutrons with energy greater than 1 MeV (see table \ref{neutron flux}).
Boulby mine laboratory has an intrinsically (cavern walls - halite) low background gamma ray activity from radionuclides contamination; the uranium and thorium content is of the order of 0.1 ppm, while the potassium concentrations is about 1130 ppm (see table \ref{primordial radionuclides}).

In the new \textsl{JIF area}, there will be a low background facility (located in the so called \textsl{Stub A})
consisting in HPGe detectors and a NaI crystal from the NaIAD experiment for crude bulk tests of activity.
Recently the crane, required to service the low background Ge detectors and ancillary equipment has been installed.  Construction of the facility clean room area recently been completed.
The Boulby Mine facility hosts a 2 kg (400 cc) germanium detector, used for material radiopurity measurements (20 cm of lead as outer shielding and 8 cm of copper as the inner shielding plus radon steel box). The setup will be sensitive to contamination of samples at the level of 0.1- 0.2 ppb.

\begin{table}
\begin{tabular}{lrrr}
\hline
  & \tablehead{1}{r}{b}{Neutron flux\\$[10^{-6}$ $n$ $cm^{-2}$ $s^{-1}]$}
  & \tablehead{1}{r}{b}{Energy}
  & \tablehead{1}{r}{b}{Ref.}   \\
\hline
LNGS       & 1.08 $\pm$ 0.02    	 & 0 - 0.05 eV	          &  \cite{Wulandari2004} \cite{Belli1989} \\
					 & 1.84 $\pm$ 0.20       & 0.05 eV - 1 keV	      &  \\ 
					 & 0.54 $\pm$ 0.01       & 1 keV	- 2.5 MeV       &  \\ 
					 & 0.27 $\pm$ 0.14       & 2.5 MeV	- 5 MeV       &  \\ 
					 & 0.05 $\pm$ 0.01       & 5 MeV	- 10 MeV	      &  \\ 
\hline
LSM        & 1.6 $\pm$ 0.1     	   & 0 - 0.63 eV	          &  \cite{chazal1998}  \\
					 & 4   $\pm$ 0.1         & 2 - 6 MeV	            &  \\
\hline 
LSC        & 3.82 $\pm$ 0.44    	         & neutrons from radioactivity	 &  \cite{Carmona2004} \\
					 & (1.73 $\pm$ 0.22  $\pm$ 0.69)  $10^{-3}$ & muon-induced neutrons in rock &	\\
\hline 
IUS \tablenote{calulated neutron flux}       & $\approx$ 2      	 & $>$ 100 keV            &  \cite{Vitaly2004} \\
					                      & $\approx$ 1        & $>$ 1 MeV	            &  \\
\hline 
SUL        & $<$ 2.7   	 & not specified          & \cite{Zdesenko1988}   \\
\hline
\end{tabular}
\caption{Neutron flux in the European underground facilities.}
\label{neutron flux}
\end{table}

\section{The ILIAS Project}

Last year a new initiative supported by the European Union (within the $6^{th}$ Framework Programme) has been started with the aim to support the European large infrastructures operating in the astroparticle physics area.
This initiative, called ILIAS (Integrated Large Infrastructures for Astroparticle Science), involes 19 European institutions (see table \ref{Europen institutions ILIAS}) and is based on 3 groups of activities: 

\begin{itemize}
	\item Networking Activities
	\subitem (N2) Deep underground science laboratories;
	\subitem (N3) Direct dark matter detection;
	\subitem (N4) Search for double beta decay;		
	\subitem (N5) Gravitational wave research;
	\subitem (N6) Theoretical astroparticle physics.
	
	\item Joint Research Activities (R\&D Projects)
	\subitem (JRA1) Low background techniques for Deep Underground Science;
	\subitem (JRA2) Double beta decay European observatory;
	\subitem (JRA3) Study of thermal noise reduction in gravitational wave detectors.
	
	\item Transnational Access Activities
	\subitem (TA1) Access to the EU Deep Underground Laboratories.
\end{itemize}

The activities involving more extensively the deep underground laboratories are N2,  JRA1 and TA1; 

The N2 network aims to implement a mechanism for the joint structuring and coordination of the four European underground facilities, in order to provide a better service to users with more efficient use of resources.

Due to the strict correlation and complementarity of the research carried on in the different laboratories, Transnational Access project to the four European deep underground laboratories is coordinated with a unique management. Within this activity the laboratories should be seen as a single infrastructure with four different installations. The ILIAS TA1 activity aims to continue and improve the experience started by LNGS within the European  $5^{th}$ Framework Programme, extending the access possibilities to the other European deep underground laboratories (LSM, LSC and IUS).

The JRA1 activity is a vast R\&D program on the improvement and implementation of innovative ultra-Low Background Techniques, and is carried out cooperatively in the four Deep European Underground Laboratories (LBT-DUSL). JRA1 should improve the quality of the services offered by underground laboratories to the scientific community in the sectors of astroparticle physics and rare-event physics, as well as improving the efficiency of the research carried on in the laboratories.
Its principal objectives are the background identification and measurement (intrinsic, induced, environmental)
and the study of the background rejection techniques (shielding, vetoes, discrimination).
There are four different Working packages:

\begin{itemize}
	\item Measurements of the backgrounds in the underground laboratories (WP1);
	\item Implementation of background MC simulation codes (WP2);
	\item Ultra-low background techniques and facilities (WP3);
	\item Radiopurity of materials and purification techniques (WP4).
\end{itemize}

\begin{table}
\begin{tabular}{lll}
\hline
  & \tablehead{1}{l}{b}{Istitution}   \\
\hline
France    	& Commissariat \`a l'Energie Atomique  \\
						& Centre National de la Recherche Scientifique \\
Italy				&	Istituto Nazionale di Fisica Nucleare \\
						&	Istituto di Fotonica e Nanotecnologie Trento \\			
						& European Gravitational Observatory \\
Germany			& Germany	Max Planck Institut für Kernphysik \\
						& Technische Universit\"at München \\
						& Max Planck Institut f\"ur Physik München \\
						& Karls Universit\"at Tubingen \\
Spain       & Zaragoza University \\
UK 					& Sheffield University \\
						& Glasgow University \\
						& London University \\
Czech Rep   & Czech Technical University in Prague \\
Denmark			& University of Southern Denmark \\
Netherland	& Leiden University \\
Finland			& University of  Jyv\"askyl\"a \\
Slovakia		& Comenius University Bratislavia \\
Greece			& Aristot University of Thessaloniki \\						
\hline
\end{tabular}
\caption{Europen institutions within the ILIAS project.}
\label{Europen institutions ILIAS}
\end{table}

Beside the four deep underground laboratories coordinated by the ILIAS project, other two deep facilities are presented in the next section: the Finland Centre for Underground Physics in Pyh\"asalmi, and the Ukranian Solotvina Underground Laboratory. In 2005 the Centre for Underground Physics in Pyh\"asalmi will reach the ILIAS community.

\section{Centre for Underground Physics in Pyh\"asalmi - CUPP}
The Laboratory is located in the mine of Pyh\"asalmi in the town of Pyh\"aj\"arvi in Finland. 
The project to host an underground laboratory in the mine was started in 1993 \cite{CUPP2001}.
The facility consists of several halls at different depths. The natural temperature increases with depth: 6 $^{0}C$ at a depth of 100 m, 16 $^{0}C$ at a depth of 1000 m (increment of about 1 $^{0}C$ per 100 m). The relative humidity in non-vantilated halls is near 100\%.
The maximum depth is 1440 m that with an average density of 2.8 $g \cdot cm^{-3}$ corresponds to about 4000 mwe. The radon concentration in air is less than 30 $Bq \cdot m^{-3}$ with ventilation.
The old part of the mine is no longer in use for mining, and will be plenty of free space to host and storage experiments.
The new mine started to operate in July 2001, the largest cavern that can be easily constructed is 20 m high, 15 m wide and 100 m long. A preliminary study, including some background measurement and rock analysis has been made (see table \ref{primordial radionuclides}) \cite{CUPP-07-2000}  \cite{Abrurashitov2003}. The headquarters are in the Oulu University.

\begin{figure}
  \includegraphics[height=.3\textheight]{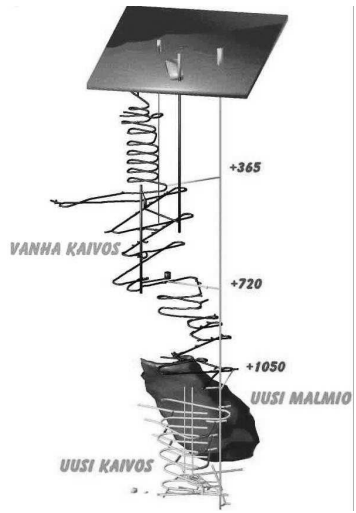}
  \caption{The Centre for Underground Physics in Pyh\"asalmi}
\label{CUPP}
\end{figure}

\section{The Solotvina Underground Laboratory - SUL}
The laboratory  was constructed in 1984 by the Institute for Nuclear Research of the Ukrainian National Academy of Sciences. It is situated on the west of Ukraine, in a small town (Solotvina) in carpathian region just near the border with Romania.
This facility is built in the operational salt (NaCl) mine 430 m underground (146 m below the sea level). It consist of one big hall (8 m high, 20 m wide and about 30 m long) and four small halls (3 m high, 3 m wide and about 3 m long) for a total area near to 1000 $m^{2}$. The natural temperature is 22-24 $^{0}C$. The Laboratory is equipped with uninterruptible power supply in order to continue the measurements if mine power system fails. 
At a depth of 430 m (that correspond to about 1000 mwe) the cosmic ray flux is reduced by a factor of about 100, at a level of 62 muons per square meter per hour. The neutron flux is less than $2.7 \cdot 10^{-6}$ $n$ $cm^{-2}$ $s^{-1}$, and radon concentration in air is less than 33 $Bq \cdot m^{-3}$. Due to a low radioactive contamination of salt, the natural gamma-rays background in the SUL is 10-100 times lower than in other underground laboratories escavated in the rock.\\
The main scientific activities at the Solotvina laboratory are concentrated in the $\beta \beta$ decay search sector \cite{Zdesenko1988}. 

\begin{figure}
  \includegraphics[height=.2\textheight]{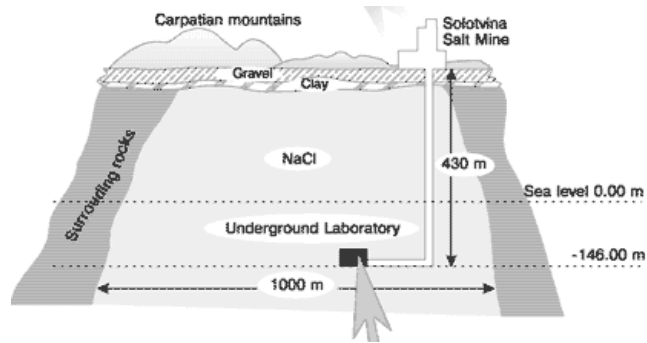}
  \caption{The Solotvina Underground Laboratory}
\label{SUL}
\end{figure}

\begin{theacknowledgments}
  This work was supported by the Italian Institute of Nuclear Physics (INFN) and by the European Union ($6^{th}$ Framework Programme) in the context of the ILIAS project (contract number UE RII3-CT2004-506222). I would like to thank Nicola Ferrari (INFN-LNGS) for useful comments.
\end{theacknowledgments}


\bibliography{sample}

\IfFileExists{\jobname.bbl}{}
 {\typeout{}
  \typeout{******************************************}
  \typeout{** Please run "bibtex \jobname" to optain}
  \typeout{** the bibliography and then re-run LaTeX}
  \typeout{** twice to fix the references!}
  \typeout{******************************************}
  \typeout{}
 }

\end{document}